\documentclass[%
 reprint,
 nofootinbib,
 amsmath,amssymb,
 aps,
 showkeys,
]{revtex4-2}

\usepackage[SIunits]{RevTeX_macro_commands}
\usepackage{local_commands}

\begin{document}

\preprint{APS/123-QED}

\title{Elementary considerations on gravitational waves from hyperbolic encounters}

\author{Martin Teuscher}
\email{teuscher@lpsc.in2p3.fr}
 \affiliation{Laboratoire de Physique Subatomique et de Cosmologie\char`,{} Univ. Grenoble-Alpes\char`,{} CNRS/IN2P3 \\ 53 avenue des Martyrs\char`,{} 38026 Grenoble Cedex\char`,{} France}
 \affiliation{Ecole Normale Supérieure de Paris \\
45 rue d'Ulm\char`,{} 75005 Paris\char`,{} France}
\author{Aurélien Barrau}%
 \email{barrau@lpsc.in2p3.fr}
\author{Killian Martineau}
\email{martineau@lpsc.in2p3.fr}
\affiliation{Laboratoire de Physique Subatomique et de Cosmologie\char`,{} Univ. Grenoble-Alpes\char`,{} CNRS/IN2P3 \\ 53 avenue des Martyrs\char`,{} 38026 Grenoble Cedex\char`,{} France}


\date{February 2024}


\begin{abstract}
We examine the main properties of gravitational waves (GWs) emitted by transient hyperbolic encounters of black holes. We begin by building the set of basic variables most relevant to setting our problem. After exposing the ranges of masses and eccentricities accessible at a given GW frequency, we analyze the dependence of the gravitational strain on those parameters and determine the trajectories resulting in the most sizeable strains. Some non-trivial behaviors are unveiled, showing that highly eccentric events can be more easily detectable than parabolic ones. In particular, we underline the correct way to extend formulas from hyperbolic to parabolic orbits. Our reasonings are as general as possible, and we make a point of explaining our considerations pedagogically. The majority of the work is based on Newtonian dynamics and aims at being a benchmark to which more accurate calculations can be compared.
\end{abstract}

\maketitle

\tableofcontents

\section{Introduction}
\label{sec:intro}
Although their status has long been debated (see, e.g., \cite{Gomes:2023xda} for a recent review), gravitational waves (GWs) are a firm prediction of general relativity (GR). Remarkably, they have recently acquired an observational status in the $\nano\hertz$ (see, e.g., \cite{NANOGrav:2023gor}) and $\unit{10^1-10^2}{\hertz}$ bands (see, e.g., \cite{LIGOScientific:2020ibl,LIGOScientific:2021djp}). The exciting prospect that measurements could be soon performed at higher frequencies has garnered substantial attention (see, e.g., \cite{Aggarwal:2020olq,Berlin:2021txa}) in the last years.\\


In this work, we specifically focus on the study of gravitational waves produced by transient hyperbolic encounters. Although a considerable amount of works (see,  e.g., \cite{Maggiore:2018sht} and references therein) are devoted to GWs from stellar collapses, exploding supernovae, compact binaries involving neutron star and/or black hole mergers -- not to mention cosmological sources, the literature is quite scarce on open trajectories. It is somewhat surprising, considering that they can naturally be expected to be very prevalent. As it is well-known, two bodies interacting gravitationally and starting from an unbounded state will always follow an open trajectory -- as long as no energy dissipative process takes over. \\

Recently, a tremendous amount of works have been devoted to scattering amplitudes (see, {\it e.g.}, \cite{Bern:2022wqg,Bjerrum-Bohr:2022blt} and references therein) for post-Minkowskian calculations (see also, e.g., \cite{Bini:2020hmy,Bini:2019nra,Bini:2020wpo,Bini:2020nsb,Damour:2019lcq,Damour:2016gwp,Damour:2017zjx,Bini:2018zxp,KoemansCollado:2019ggb,Cheung:2018wkq,Kosower:2018adc,Bern:2019nnu,Bautista:2019tdr,KoemansCollado:2019ggb,Cristofoli:2019neg} and references therein). It is quite remarkable that, using relativistic quantum field theory, it seems possible to use modern amplitude techniques to derive classical post-Minkowskian expansions of general relativity -- supporting the idea that Einstein's gravity is suitably defined perturbatively from a path-integral loop expansion. This framework relies only on developments based on the gravitational constant, allowing the analysis to be carried out at velocities close to that of light. It is also, by construction, ideally suited for open trajectories. Obviously, a detailed comparison between our results and the more refined calculations quoted above is beyond the scope of this work. Our goal here is to thoroughly investigate a simple approximation so as to provide the community with a useful benchmark. To the best of our knowledge, the purely Newtonian approach (supplemented with GW emission) has not been yet fully unveiled and explored as far as optimization is concerned. As the situation is already quite involved at this level, it seems important to untangle the classical subtleties that might help understanding the full picture in a more accurate treatment.

It is also worth mentioning that remarkable analytical results were obtained on dynamical captures in black hole binaries (see  e.g. \cite{East:2012xq,Nagar:2020xsk,Gamba:2021gap}). In this study we only consider hyperbolic trajectories without capture, yet the framework could be useful to consider in the future.\\

The aim of this article is to analyze, in full generality, how the geometry of these unbounded trajectories leaves imprints in the generated GW strain and to provide a clearer understanding of the qualitative picture. We will show that determining the trajectory resulting in the largest emitted strain, referred to as the ``optimal" trajectory (for a given detection frequency), is far from trivial. In a nutshell, the optimal situation occurs when masses reach the maximal value allowed by physical constraints and follow a parabola (that is, the trajectory with the smallest possible eccentricity $e=1$). If, however, masses are fixed at a given value lower than this upper limit, the most favorable trajectory turns out to be -- somewhat counter-intuitively -- the one with the {\it largest} possible eccentricity.\\


Whevener it is possible our statements are derived with the fewest possible assumptions, thus are often of full generality; we also remain as pedagogical as possible. In section \ref{sec:notations} we introduce all relevant physical quantities, recall how to compute the GW strain emitted on an open orbit and address some moot points, especially about parabolic orbits. Section \ref{sec:main} constitutes the core of this work, in which the aforementioned result is proven and more generally the parameter space of our problem is investigated.\\

Throughout all this work we use natural units where $\hbar = c = k_B = 1$, occasionally reintroducing these constants for the sake of clarity.

\section{Generalities about hyperbolic encounters}
\label{sec:notations}

\subsection{Preliminaries}
\label{sub:notations}

\begin{figure}
    \centering
    \includegraphics[width=0.48\textwidth]{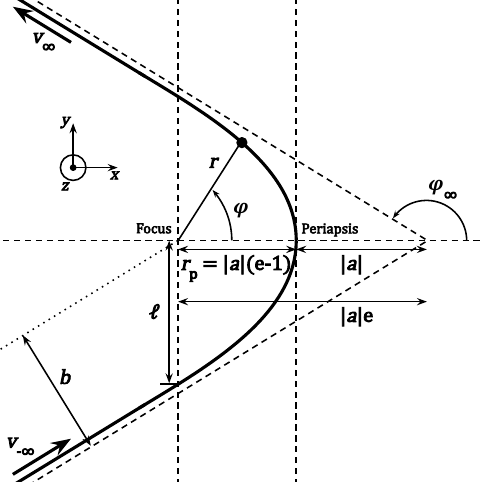}
    \caption{Geometrical parameters describing an hyperbolic trajectory in the center-of-mass frame. Notice that with such conventions, $\ihp$ grows strictly from $-\ihp_\infty$ to $\ihp_\infty$ anticlockwise, and $\ihp_p = 0$ at periapsis.}
    \label{fig:hyperbola}
\end{figure}

Let us consider a hyperbolic gravitational encounter between two bodies of masses $m_1$ and $m_2$. In practice, we shall focus on black holes (BHs) as they are the most susceptible to be observed. We denote $M\equiv m_1+m_2$ their total mass, $\mu\equiv \flatfrac{m_1 m_2}{M} \leqslant \flatfrac{M}{4}$ their reduced mass, and $\vb{r}_1 = \frac{\mu}{m_1} \vb{r},\ \vb{r}_2 = -\frac{\mu}{m_2} \vb{r}$ their respective positions, with $\vb{r}=\vb{r}_1 - \vb{r}_2$ the separation vector in the center-of-mass frame. 
We use polar coordinates $(r,\ihp)$ and define $\w\equiv \dv*{\ihp}{t}$. $\vb{r}$ is assumed to describe an hyperbola, whose geometrical parameters are depicted on \figref{fig:hyperbola}. In Newtonian dynamics, introducing $\kappa \equiv \G M$, 
the links between {\it geometrical} parameters $b$ (impact parameter, or semi-minor axis), $e$ (eccentricity), $\ell$ (semi-latus rectum), $a$ (semi-major axis), $\ihpinf$ (outgoing angle) and $r_p$ (periapsis radius, or distance to focus at closest approach), and {\it dynamical} parameters $\vinf$ (excess velocity), $\w\equiv \dv*{\ihp}{t}$ (pulsation), $\w_p \equiv \w|_{r=r_p}$ (pulsation at periapsis) and $\cal{E}$ (conserved energy), are as follows:
\begin{subequations}
    \label{eq:hyperbola-defs}
    \begin{align}
    \label{eq:def-polarcoo}
     &   r(\ihp) = \frac{\ell}{1+e\cos\ihp}\, ,\qquad -\ihpinf < \ihp < \ihpinf \\
    \label{eq:def-phiinf}   
     &   \ihpinf = \arccos(-\frac{1}{e})\in\interval{]}{\frac{\pi}{2}}{\pi}{]} \\
    \label{eq:def-eccentricity}
     &   e = \sqrt{1+\frac{b^2}{a^2}} > 1 \\
    \label{eq:def-semilatusrectum}
     &   \ell = -a(e^2-1) = \left(\frac{\kappa}{\w_p^2}\right)^{1/3} (e+1)^{4/3}\\
    \label{eq:def-semimajoraxis}
     &   a = - \frac{\kappa}{\vinf^2} = - \left(\frac{\kappa}{\w_p^2}\right)^{1/3} \frac{(e+1)^{1/3}}{e-1} < 0 \\
    \label{eq:def-impactparam}
     &   b = \left(\frac{\kappa}{\w_p^2}\right)^{1/3} \frac{(e+1)^{5/6}}{(e-1)^{1/2}} \\
    \label{eq:def-v_inf}
      &  \vinf = (\kappa\w_p)^{1/3} \frac{(e-1)^{1/2}}{(e+1)^{1/6}} \\
    \label{eq:def-omega}
      &  \w(\ihp) = \frac{b\vinf}{r^2(\ihp)} = \w_p\left(\frac{1+e\cos\ihp}{1+e}\right)^2 < \w_p\\
    \label{eq:def-periapsisdistance}
     &   r_p = \frac{\ell}{1+e} = \left(\frac{\kappa}{\w_p^2}\right)^{1/3}(e+1)^{1/3} < b \\
    \label{eq:def-energy}
      &  \cal{E} = -\frac{\kappa \mu}{2 a} = \frac{\G m_1 m_2}{2\ell}(e^2-1) > 0 \mperiod
    \end{align}
\end{subequations}
The reason we emphasize the somewhat unusual variable $\w_p$ in the previous equations is explained in section \ref{sub:paraboliclimit} and its role will become crucial in section \ref{sub:parameter-space}.

\smallskip
In the post-Newtonian limit of GR, it is assumed that the massive bodies still follow the hyperbolic Newtonian trajectory, thus Eqs.~\eqref{eq:hyperbola-defs} remain valid. However, at next-to-leading order the moving bodies will source gravitational waves because of their changing quadrupolar distribution. The lowest multipole moments govern the emission of GWs by non-relativistic sources. The quadrupole formula is obviously an approximation, however known to work with great accuracy \cite{Blanchet:2019zlt}. It is suitable for the system considered in this work -- the trajectory does not need to be bounded. Post-Minkowskian approximation will not be considered in this work but should obviously be studied in the future.

At lowest order in the multipole expansion, the metric perturbation tensor $h_{\mu\nu}$ in the  transverse-traceless (TT) gauge can be written as (see \cite{Maggiore:2007ulw})
\begin{equation}
    h_{ij}^{\TT}(t+R/c,\vb{R} = R\vu{m}) = \Lambda_{ij}^{kl} \frac{2\G}{R}\ddot{I}_{kl}(t)\mcomma
\end{equation}
where \vspace{-0.4cm}
\begin{equation}
\label{eq:def-quadrupole}
    I_{ij} = \mu r_i r_j
\end{equation}
is the quadrupole moment of Newtonian binaries, $R$ is the distance to the observer, much greater than $r$ the source's typical size, and $\Lambda_{ij}^{kl}$ is a projection tensor defined, for a wave propagating in the $\vu{m}$ direction, as
\begin{equation}
    \Lambda_{ij}^{kl} \equiv P_i^k P_j^l - \pref{2}P_{ij}P^{kl},\qquad P_{ij} \equiv \delta_{ij}-m_i m_j \mperiod
\end{equation}
Furthermore, if $(\vu{u},\vu{v})$ is a basis of the plane transverse to $\vu{m}$ such that $(\vu{m},\vu{u},\vu{v})$ direct, the plus- and cross-polarization amplitudes $h_+, h_\times$ can be obtained as 
\begin{equation}
    h_{+,\times} = \pref{2} e_{+,\times}^{ij} h_{ij}^{\TT},\qquad \begin{array}{rcl}
     e_+^{ij} &\equiv& u^i u^j - v^i v^j\\[0.1cm]
     e_\times^{ij} &\equiv& u^i v^j + v^i u^j \mperiod
    \end{array}
\end{equation}
We also define the trace-free multipole moment 
\begin{equation}
    Q_{ij} \equiv I_{ij} - \pref{3}I_k^k \delta_{ij} \mcomma
\end{equation}
from which the power emitted in GWs by the binary system can be expressed as
\begin{equation}
    P\equiv \frac{\G}{5}\left\bra\dddot{Q}_{ij}\dddot{Q}^{ij}\right\ket \mcomma
\end{equation}
where $\bra\cdots\ket$ denotes an average over a few periods of the GWs; subtleties about the definition of a burst's period are developed in section \ref{ssub:GWfreq}. 

Since $\Lambda_{ij}^{kl}\delta_{kl} = 0$, the perturbations in the TT gauge can also be expressed as:
\begin{equation}
    h_{ij}^{\TT}(t+R/c,\vb{R} = R\vu{m}) = \Lambda_{ij}^{kl} \frac{2\G}{R}\ddot{Q}_{kl}(t) \mperiod
\end{equation}

Finally, we define a typical amplitude for the strain by 
\begin{equation}
    \mathfrak{h} \equiv \sqrt{h_+^2 + h_\times^2} = \sqrt{\pref{2}h_{ij}^{\TT}h^{\TT\, ij}} \mperiod
\end{equation}

To simplify the analysis in what follows, a wave emitted perpendicular to the plane of the hyperbola is considered, with $\vu{m}=\vu{z}$ and $\vu{u}, \vu{v} = \vu{x}, \vu{y}$.

\subsection{Main features of the GW signal}
\label{sub:def-h-f-and-t}

\subsubsection{Strain}
\label{ssub:GWstrain}

\begin{figure}
    \centering
    \includegraphics[width = 0.48\textwidth]{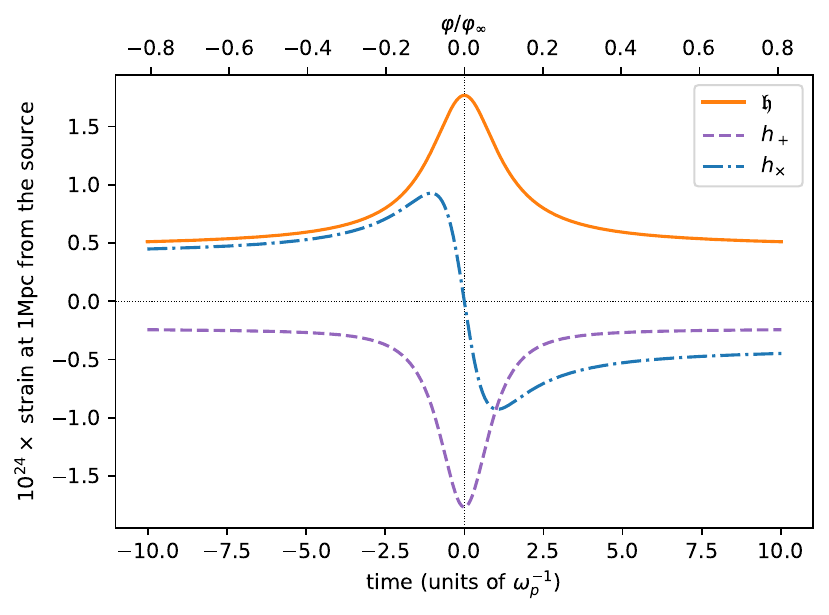}
    \caption{Plus-polarization, cross-polarization and typical strain for an unbounded event involving two equal mass BHs of $m_{1,2} = \unit{3\times 10^{-5}}{\solarmass}$ such that $\w_p = \unit{1}{\giga\hertz}$ and $e=2$. Far from $\ihp=0$ the apparent linear relation between time and angle breaks down, c.f. Eqs.~\eqref{eq:tandphi-far}--\eqref{eq:tandphi-close} later in the text.}
    \label{fig:hc-hplus-hcross}
\end{figure}

From \eqref{eq:def-quadrupole} and previous conventions one has
\begin{equation}
    Q_{ij} = \pref{3}\mu r^2(\ihp) \begin{pmatrix}
        3\cos^2\ihp-1 & 3\cos\ihp\sin\ihp & 0 \\
        3\cos\ihp\sin\ihp & 3\sin^2\ihp -1 & 0 \\
        0&0&-1
    \end{pmatrix}
\end{equation}
which yields (for an efficient calculation that avoids computing $\Lambda_{ij}^{kl}$, Eqs.~(3.65) and (3.66) of \cite{Maggiore:2007ulw} can be used)
\begin{widetext}
\begin{align}
\label{eq:hplus-value}
    h_+ &= - \frac{\G \mu (\kappa\w_p)^{2/3}}{Rc^4 (e+1)^{4/3}}\left[2e^2 +5e\cos{\ihp} +4\cos{2\ihp} +e\cos{3\ihp}\right] \\
\label{eq:hcross-value}
    h_\cross &= -\frac{2\G \mu (\kappa\w_p)^{2/3}}{Rc^4 (e+1)^{4/3}}\sin{\ihp}\left[3e +4\cos{\ihp} +e\cos{2\ihp}\right]\\
    \label{eq:hc-value}
    \mathfrak{h} &= \frac{\sqrt{2}\G \mu (\kappa\w_p)^{2/3}}{Rc^4 (e+1)^{4/3}}\left[8 + 13 e^2 + 2 e^4 + 2 e (12 + 5 e^2) \cos{\ihp} + 13 e^2 \cos{2\ihp} + 2 e^3 \cos{3\ihp}\right]^{1/2}\\
     \label{eq:power-value}
    P &= \frac{4\G\mu^2(\kappa\w_p)^{4/3}}{15c^5(e+1)^{8/3}}\w^2(\ihp)\left[24+13e^2+48e\cos{\ihp} +11e^2\cos{2\ihp}\right] \mcomma
\end{align}
\end{widetext}
where we have reintroduced the speed of light for clarity. These expressions are plotted on \figref{fig:hc-hplus-hcross}. Notice that $P$ vanishes when $\ihp\to\pm\ihp_\infty$ -- no power is emitted outside the burst, but that the oddness of the \textit{sin} function implies $h_\cross(\ihp\to -\ihp_\infty) \neq h_\cross(\ihp\to +\ihp_\infty)$: the cross-polarization exhibits a well-known linear memory effect \cite{Favata_2010}.

From \eqref{eq:hc-value} and \eqref{eq:power-value} it can be seen that, consistent with intuition, both $\mathfrak{h}$ and $P$ are maximal at the periapsis, where the trajectory undergoes the strongest deviation. At this point, one finds (hereafter a $p$ index denotes a quantity evaluated at the periapsis):
\begin{subequations}
    \begin{align}
    \label{eq:def-hc}
        \mathfrak{h}\subsc{max} &= \mathfrak{h}_{p} = \frac{2\G \mu (\kappa\w_p)^{2/3}}{Rc^4}\frac{e+2}{(e+1)^{1/3}} \\
        P\madmax &= P_p = \frac{32\G\mu^2 \kappa^{4/3} \w_p^{10/3}}{5c^5(e+1)^{2/3}} \mperiod
    \end{align}
\end{subequations}

\subsubsection{Frequency}
\label{ssub:GWfreq}

The unbounded trajectory being aperiodic by essence, the produced GWs do not exhibit a well-defined frequency $f$, at least not as commonly understood for elliptic orbits. To bypass this problem, an effective frequency is defined as the one corresponding to the peak of the signal Fourier transform, or simply as $\w(\ihp)\equiv\dv*{\ihp}{t}$ evaluated at closest approach, i.e. $2\pi f\equiv\w_p$. Both definitions are fortunately equivalent \cite{caldarola2023effects}.
In addition, by massaging \eqref{eq:hyperbola-defs}, one obtains
\begin{equation}
\label{eq:pseudo-keplers-law}
    \w_p^2 r_p^3 = \G M(e+1) \mperiod
\end{equation}
We recover an analog of Kepler's law for unbounded orbits, which confirms the natural interpretation of $\w_p$ as the wave's pulsation during the burst.

\subsubsection{Duration and bandwidth}
\label{ssub:GWtime}

It is also crucial to investigate the time spent by the signal in a given frequency band, as this information is paramount for accurately evaluating the sensitivity of a detector to an incoming wave \cite{Moore:2014lga}. Similarly to the work done in \cite{Garc_a_Bellido_2017}, by invoking the conservation of momentum $r^2\dot{\ihp}=\text{constant} = b\vinf$, it follows:
\begin{align} 
t(\ihp)& = \pref{b\vinf} \int \frac{\ell^2\dd{\ihp}}{(1+e\cos\ihp)^2} \\
   \label{eq:time-expression}
   \begin{split}
      & =\frac{1}{\w_p}\frac{e+1}{e-1}\left[\frac{e\sin\ihp}{1+e\cos\ihp}\right. \\
      &\qquad\quad\left. -\frac{2}{\sqrt{e^2-1}}\arctanh\left(\sqrt{\frac{e-1}{e+1}}\tan\frac{\ihp}{2}\right)\right] \mcomma
   \end{split}
\end{align}
where the origin of time has been defined in such a way that $t_p = t(\ihp_p)=0$. Inverting \eqref{eq:def-omega} allows to replace $t(\ihp)$ by $t(\w)$ (with two branches), which when expanded near $\w\approx\w_p$ provides the time $\Delta t\equiv t_+(\w_p - \Delta\w)-t_-(\w_p - \Delta\w)$ spent by the system in the frequency band $\interval{[}{\w_p-\Delta\w}{\w_p}{]}$:
\begin{equation}
\label{eq:deltat-approx}
    \Delta t \approx \frac{2}{\w_p}\sqrt{1+\frac{1}{e}}\sqrt{\frac{\Delta\w}{\w_p}}
\end{equation}
with $\Delta\w \ll\w_p$. We emphasize that during this time the signal does not go ``through" a bandwidth centered around $\w_p$, since $\forall \ihp,\ \w(\ihp)\leqslant \w_p$. It rather drifts into this bandwidth from lower frequencies before drifting back, reaching its maximum frequency $\w_p$ in the middle of its course.



\subsection{The parabolic limit}
\label{sub:paraboliclimit}

To conclude this section, we expose how to properly take the parabolic limit $e \to 1$ in all the precedent expressions. This issue is often disregarded in the literature despite its utterly non-trivial nature.

\smallskip
In many other articles \cite{Garc_a_Bellido_2018, caldarola2023effects} one can read expressions such as
\begin{subequations}
\label{eq:wrongvariables}
\begin{align}
    \w_p &= \frac{\vinf}{b}\frac{e+1}{e-1} \\
    \label{eq:wrongvariables-strain}
    \mathfrak{h}_{p} &=  \frac{2\G \mu \vinf^2}{Rc^4}\frac{e+2}{e-1} \\
     P_p &= \frac{32G\mu^2 \vinf^6}{5c^5 b^2}\frac{(e+1)^2}{(e-1)^4} \mperiod
\end{align}
\end{subequations}
which are consistent with \eqref{eq:hyperbola-defs}, but rather use $(\kappa,\vinf, b)$ as basic variables. Here, the value $e=1$ blatantly induces divergences, which are artifacts arising from the simultaneity of $e=1$, $\vinf = 0$ (since $\cal{E}=0$), $b=\infty$ and $a=\infty$ on a parabola. Finding finite, non-zero limits for \eqref{eq:wrongvariables} requires to untangle the interdependence of these parameters, by imposing that some quantities must remain fixed. First, $M$ is kept constant so only the geometry of the trajectory matters but not physical objects. Two equivalent possibilities then exist:
\begin{enumerate}
    \item {\it Finite semi-latus rectum $\ell$.} The polar equation \eqref{eq:def-polarcoo}, from which all the dynamics can be derived, does not lead to any divergence when $e=1$. Hence, considering $\ell$ as constant in this expression when $e\to 1$ is safe.
    \item {\it Finite periapsis radius $r_p$.} \figref{fig:hyperbola} provides the geometrical intuition that one can fix the distance $r_p$ between the periapsis and the focus of the trajectory, then evolve the hyperbola into a parabola from this point. However, when $e\to 1$ Eq.~\eqref{eq:def-periapsisdistance} shows that $r_p = \ell /2$, hence this scenario boils down to holding $\ell$ fixed as well, and the previous case is recovered.
\end{enumerate}
 Plugging back this condition in \eqref{eq:def-semilatusrectum}  et seq. constrains the limits for the aforementioned parameters as follows:
    \begin{subequations}
        \label{eq:parabolic-limit}
    \begin{align}
        a&\limsim_{e\to 1} -\frac{\ell}{2(e-1)} \\
        b &\limsim_{e\to 1} \frac{\ell}{\sqrt{2}\sqrt{e-1}} \\
        \vinf^2 &\limsim_{e\to 1} \frac{2\kappa}{\ell}(e-1) \mperiod
    \end{align}
 \end{subequations}
However, when $e=1$, $\text{\eqref{eq:def-periapsisdistance}} \implies r_p^3 \w_p^2 = 2 \kappa$, therefore $\ell\underset{e\to 1}{=} \const \iff r_p\underset{e\to 1}{=} \const \iff\w_p\underset{e\to 1}{=} \const$. Thus, \eqref{eq:parabolic-limit} can alternatively be rewritten with $\w_p$, as
\begin{subequations}
        \label{eq:parabolic-limit-omega}
    \begin{align}
        a(e-1) &\limsim_{e\to 1} \const = - \left(\frac{2\kappa}{\w_p^2}\right)^{1/3} \\
        b\sqrt{e-1} &\limsim_{e\to 1} \const = \sqrt{2}\left(\frac{2\kappa}{\w_p^2}\right)^{1/3} \\
        \frac{\vinf^2}{e-1} &\limsim_{e\to 1} \const = \pref{2^{1/3}} (\kappa\w_p)^{2/3} \mperiod
    \end{align}
 \end{subequations}
A bit of algebra shows that inserting \eqref{eq:parabolic-limit-omega} into \eqref{eq:wrongvariables} eliminates all $e-1$ factors, therefore removing all spurious divergences. 

This sheds light on the peculiar choice of $(\kappa, e, \w_p)$ triplet as basic variables we made in \eqref{eq:hyperbola-defs}: the troublesome elimination of fake parabolic divergences can be avoided, since substituting $e=1$ in Eqs.~\eqref{eq:hplus-value}--\eqref{eq:power-value} becomes effortless; and, simultaneously, the frequency of the emitted GW is immediately apparent.

\section{Exploring the parameter space}
\label{sec:main}

This section examines in details which part of the hyperbolic encounter parameter space can generate observable GW strains. Points of this space contributing to the most sizeable strains receive a particular attention. Eventually, the assumptions required to derive the results of this section are summarized and discussed at the end.


\subsection{Physical bounds}
\label{sub:parameter-space}

A GW burst of frequency $\w_p$ -- as defined in section \ref{ssub:GWfreq} -- will trigger a response from a detector providing $\w_p$ lies within the bandwidth of this instrument.\footnote{In this work, anything likely to change the GW frequency between emission and detection is ignored, e.g. redshift $z$ or boosts with Lorentz factor $\gamma$. They can nonetheless be included without much trouble, as it is enough to replace $\w_p$ by $(1+z)^{-1}\w_p$ or $\gamma^{-1} \w_p$ in the entire analysis.} From an experimental point of view, it is therefore well motivated to pursue our analysis at {\it fixed} $\w_p$: all trajectories resulting in $\w_p \neq \w_D$ are put aside, where $\w_D$ is a given frequency inside some detector's bandwidth.



Moreover, as can be seen from \eqref{eq:hyperbola-defs}, three parameters are enough to fully constrain the dynamics of the encounter,\footnote{This is because the trajectory is only sensible to the {\it total} mass $M$. Quantities relying also on the reduced mass $\mu$ like $\mathfrak{h}$ depend on four parameters.} thus fixing $\w_p$ leaves two degrees of freedom.  The most relevant choice is to span them with $\kappa=\G M$ and $e$. Not only was it the outcome of the discussion in section \ref{sub:paraboliclimit}, but also the distribution of $\kappa$ may be constrained by astrophysical or cosmological data (see \cite{Byrnes:2021jka} and references therein), while using $e$ eases computations. 

\smallskip
Besides, some regions of this 2D parameter space are forbidden by physical constraints:
\begin{enumerate}[label={\it (\roman*)}]
    \item \label{it:nomerger} {\it No-merger} constraint: any dynamics resulting in a BH merger contradicts the original assumption that the trajectory is an open hyperbola, hence must be ruled out from this analysis. It shall therefore be imposed $r_p > R_{s1}+R_{s2} \equiv R_s = \flatfrac{2\kappa}{c^2}$, the Schwarzschild radius.
    \item  \label{it:speed limit} {\it Speed limit} constraint: a trajectory where the relative velocity of the bodies exceeds the speed of light is unphysical, so one shall also set the constraint $\vmax < c$, with $\vmax = v_p = r_p \w_p$ attained at the periapsis.
\end{enumerate}

It should be noted that handling trajectories (almost) saturating inequality \ref{it:speed limit} is dangerous because these break the post-Newtonian expansion, so self-consistency may require to replace this bound by $\vmax < \eps c$, $\eps\ll 1$. Translating the former bound $\vmax < c$ into a stronger bound on $\vinf$, based on Newtonian laws, as exposed in \cite{Garc_a_Bellido_2018}, can be questionable: even if $\vinf\to c$, $\vmax$ will in reality never exceed $c$, conflicting with Newtonian dynamics predictions. One can therefore easily obtain safer limits by taking $\eps$ (much) smaller that unity. We would however like to emphasize that if relativistic corrections will inevitably change the amplitude of the generated strain, upper bounds on detectable masses and what derives from them is valid in a general context, as we demonstrate in the next paragraph. Hence, the main picture is very likely to remain correct.

\bigskip
It is nevertheless possible to derive a property that remains valid even for highly relativistic trajectories. First the relationship $\vmax = v_p = \w_p r_p$ consistently holds -- as it is purely a geometrical definition. Equating $\w_p$ to $\w_D$ the detector's frequency, \ref{it:nomerger} implies $\vmax > R_s \w_D$ and\footnote{Relating the Schwarzschild radius and the speed-of-light limit is actually quite intuitive. In Newtonian gravity, a massive body on a circular trajectory of radius $R_s$ travels at speed $c$.} \ref{it:speed limit} further yields $c > R_s \w_D$, i.e.
\begin{align}
\label{eq:Mcritical}
     M <\ & \frac{c^3}{2\G \w_D} \\
     &=\unit{1,0\times 10^5}{\solarmass}\left(\frac{\unit{1}{\hertz}}{\w_D}\right) \mperiod
\end{align}
Eq.~\eqref{eq:Mcritical}, stating the existence of an upper bound on the BH masses observable by {\it any} GW detector, is actually very general: it is a non-perturbative result true outside the post-Newtonian expansion, and for both hyperbolic and elliptic trajectories.\\

\begin{figure*}
    \centering
\includegraphics[width=0.48\textwidth]{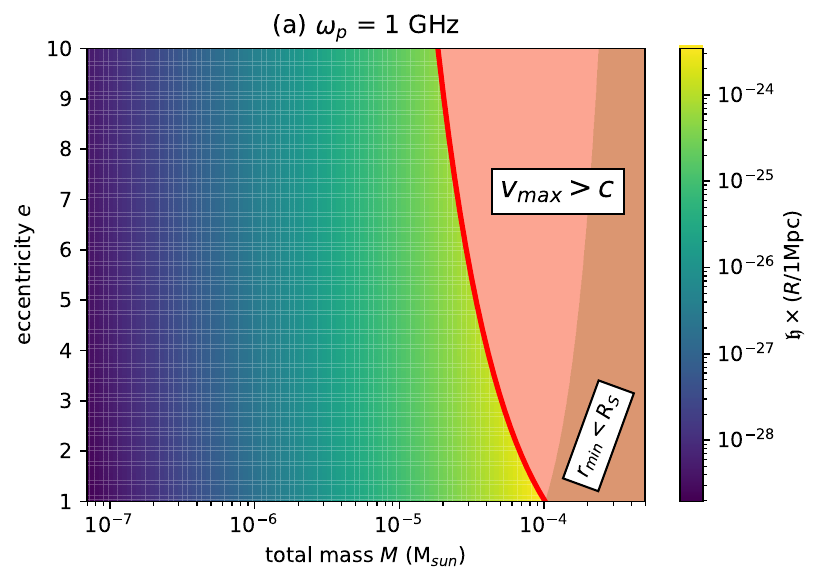}
\includegraphics[width=0.48\textwidth]{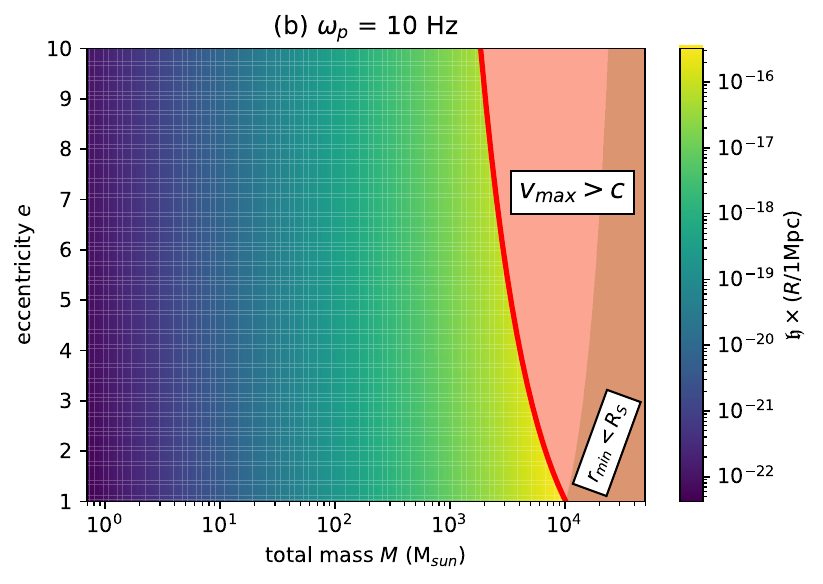}
    \caption{Typical strain $\mathfrak{h}$, normalized at a distance of $\unit{1}{\mega\parsec}$, as a function of eccentricity and total mass, for two values of $\w_p$ and assuming $m_1=m_2$. When one variable is fixed, one has $\mathfrak{h} \propto M^{5/3}$ and $\mathfrak{h} \propto e^{2/3}$ (for $e$ sufficiently large). The solid red line indicates the saturation of inequality \eqref{eq:Mcritical-newtonian} and delimits the physically allowed region of the parameter space. As is rigorously proven in section \ref{sub:abs-max-strain}, the intersection of this line with the line $e=1$ is the point where $\mathfrak{h}$ is maximal.}
    \label{fig:2Dparam}
\end{figure*}

We henceforth come back to the Newtonian description, which by plugging \eqref{eq:hyperbola-defs} inside \ref{it:speed limit} allows to refine \eqref{eq:Mcritical} into:
\begin{equation}
\label{eq:Mcritical-newtonian}
    M < \frac{c^3}{\G \w_D(e+1)}\mperiod
\end{equation}
The maximal observable mass is therefore detectable at $\w_D$ when $e=1$, i.e. when it follows a parabolic orbit. Bound \eqref{eq:Mcritical-newtonian} as well as values for $\mathfrak{h}(\kappa, e, \w_p)$ are presented on \figref{fig:2Dparam}. It is observed that $\mathfrak{h}$ is a growing function of both $e$ and $M$ when the other variable remains constant: $\mathfrak{h} \propto M^{5/3}$ and  $\mathfrak{h} \propto e^{2/3}$ for $e\gg 1$. As appears on the same figure, some algebra shows that in this framework the region excluded by \ref{it:speed limit} includes the one forbidden by \ref{it:nomerger}. 

In the next two subsections, we explain how to rigorously locate the point on the $(M,e)$-plane where $\mathfrak{h}$ is maximal. As \figref{fig:2Dparam} exhibits, it is the intersection of the physical region boundary with the line $e=1$. Since $\mathfrak{h}\propto \mu$, we limit our attention to cases where $\mu$ is maximal for a given $M$, that is when $m_1=m_2$ and $\mu=\flatfrac{M}{4}$.


\subsection{Globally maximal strain}
\label{sub:abs-max-strain}

Finding the maximal strain at fixed $\w_p$ under constraints \ref{it:nomerger} and \ref{it:speed limit} is an optimization problem, hence we use the Karush-Kuhn-Tucker (KKT) theorem. The generic lagrangian with Lagrange mutipliers $\zeta$, $\lambda$ and $\xi$ would read
\begin{dmath}
    \cal{L}\subsc{KKT}(\kappa,e,\w_p) = \mathfrak{h}_{p}(\kappa,e,\w_p)\\+\zeta \w_p(\kappa,e,\w_p)\\ + \lambda (\vmax(\kappa,e,\w_p) - c) \\ + \xi (R_s(\kappa) - r_p(\kappa,e,\w_p)) \mcomma
\end{dmath}
but this optimization problem can be greatly simplified. First, using $\w_p$ as a variable renders the first multiplier unnecessary. Second, the third multiplier is redundant since \ref{it:speed limit}$\implies$\ref{it:nomerger}, c.f. \figref{fig:2Dparam}. The lagrangian then reduces to:
\begin{dmath}
\label{eq:KKTLag}
     \cal{L}\subsc{KKT}(\kappa,e) = \mathfrak{h}_{p}(\kappa,e) + \lambda (\vmax(\kappa,e) - c) \\
   =\frac{\kappa^{5/3}\w_p^{2/3}}{2Rc^4}\frac{e+2}{(e+1)^{1/3}}\\ \quad + \lambda\left[(\w_p\kappa(1+e))^{1/3} - c\right] \mperiod
\end{dmath}
Invoking the KKT conditions, if the constraint is not saturated at the optimum $(\kappa_*,e_*)$ of $\cal{L}\subsc{KKT}$, then $\lambda=0$. Let us suppose this is the case. From $\pdv*{\cal{L}\subsc{KKT}}{e}=0$, we can express $\lambda(\kappa_*,e_*)$, which is zero for $e\geqslant 1$ only if $\kappa=0$. Thus, this situation minimizes $\mathfrak{h}_{p}$ rather than maximizing it. Consequently, the maximum must be reached for some $(\kappa_*,e_*)$ that saturates the constraint, i.e.
\begin{equation}
\label{eq:optimizing-constraint}
    \kappa_*(e_*+1) = \frac{c^3}{\w_p} \mperiod
\end{equation}
When plugging back \eqref{eq:optimizing-constraint} in the expression of $\mathfrak{h}_{p}$, it becomes a usual one-variable function, whose maximum is located at $e_*=1$.\footnote{Note that $\var{\cal{L}\subsc{KKT}}(\kappa_*,e_*)=0$ cannot be assumed anymore: now that the constraint is saturated, the maximum is reached on the edge of the validity domain and therefore may not be a {\it local} maximum. In fact, by inserting \eqref{eq:optimizing-constraint} 
in the expression of $\mathfrak{h}_{p}$, one can check $\pdv*{\mathfrak{h}_{p}}{e}\/ (e_*=1)\neq 0$, yet this is the correct global maximum.}  Substituting $e_*=1$ in \eqref{eq:optimizing-constraint} leads to $\dps \kappa_* = \frac{c^3}{2\w_p}$, i.e. \eqref{eq:Mcritical}. Thus, we have demonstrated that {\it the open keplerian trajectory producing the strongest GW strain at a given observational frequency $\w_p$ is a parabola, for two BHs of equal masses $m_1 = m_2 = \flatfrac{M}{2} = \flatfrac{c^3}{4\G\w_p}$ that pass next to each other at a distance equal to twice their Schwarzschild radius.} The value of the GW strain in such a scenario is
\begin{align}
    \label{eq:absolute-max-h}
    \mathfrak{h}_{p}^{\text{opt}} (R,\w_p) &= \frac{3}{4R}\frac{\kappa}{c^2} = \frac{3}{8R}\frac{c}{\w_p} \\
    \label{eq:absolute-max-h-fiducial-hertz}
    &=3,6\times 10^{-18}\left(\frac{\unit{1}{\giga\rm{pc}}}{R}\right)\left(\frac{\unit{1}{\hertz}}{\w_p}\right) \\
    \label{eq:absolute-max-h-fiducial-gigahertz}
     &=3,6\times 10^{-24}\left(\frac{\unit{1}{\mega\rm{pc}}}{R}\right)\left(\frac{\unit{1}{\giga\hertz}}{\w_p}\right) \mperiod
\end{align}
This outcome holds under the assumption that BHs follow a trajectory allowed by Newtonian dynamics, and that the GWs are given by the first order of the multipole expansion.

\subsection{Maximal strain in non-ideal scenarios}
\label{sub:suboptimal-max-strain}

The last result, although intuitive -- masses undergo a stronger deviation on a parabola -- was derived assuming it was possible to optimize the strain over BH's masses. In real-life scenarios, this is not what would happen as the masses of the incoming bodies may be suboptimal ($M < \flatfrac{c^3}{2\G\w_p}$) and lie to the left of the red curve in \figref{fig:2Dparam}. Describing more accurately the parameter space observable by GW detectors requires to address the following question: in a setup where the mass $M$ is given independently of $\w_p$, what is the trajectory maximizing the strain and which value can it reach?

To answer this, one considers again the KKT lagrangian \eqref{eq:KKTLag} which is now a function of $e$ only. The same reasoning about the Lagrange multiplier holds, meaning that Eq.~\eqref{eq:optimizing-constraint} remains valid, except that $\kappa$ is now given {\it a priori}. This implies that the eccentricity is bounded from above by
\begin{equation}
\label{eq:bound-on-e}
    e \leqslant e_* =\frac{c^3}{\kappa\w_p} - 1
\end{equation}
i.e. BHs on any orbit with a greater eccentricity cannot generate GWs of frequency $\w_p$ without exceeding $c$. The highest GW strain, now reached for an eccentricity saturating \eqref{eq:bound-on-e} which no longer relates to a parabola, reads
\begin{align}
\label{eq:general-suboptimal-h}
    \mathfrak{h}_{p}^{\text{subopt}}(R,\w_p,\kappa) &= \frac{1}{2R}\frac{\kappa}{c^2}\left(1+\frac{\kappa\w_p}{c^3}\right) \\
    \label{eq:suboptimal-h-fiducial}
    \begin{split}
    &=2,5\times 10^{-25}\left(\frac{\unit{1}{\mega\rm{pc}}}{R}\right)\left(\frac{M}{\unit{10^{-5}}{\solarmass}}\right) \\
    &\qquad \times\left(1 + \unit{0,05}{}\frac{M}{\unit{10^{-5}}{\solarmass}}\frac{\w_p}{\unit{1}{\giga\hertz}}\right) \mperiod
    \end{split}
\end{align}
In result, there exist hyperbolic trajectories leading to a stronger strain that the rather intuitive parabolic trajectory. The reason for this can be grasped through the following argument. What has to be fixed in the analysis is the {\it angular} velocity at closest approach, $\w_p$. Hence, the further away from the origin the body passes, the higher its speed must be to maintain this angular velocity. However, the strain itself is sensitive to the absolute velocity (see \eqref{eq:wrongvariables-strain}), meaning a body on a highly eccentric orbit would create a higher GW strain. Nonetheless, if the parabolic orbit leads to the body reaching the speed of light regardless ($r_p = R_s$), then the parabola becomes the optimal trajectory, and the strain is the highest amongst all trajectories.

\begin{figure*}
    \centering
    \includegraphics[width=\textwidth]{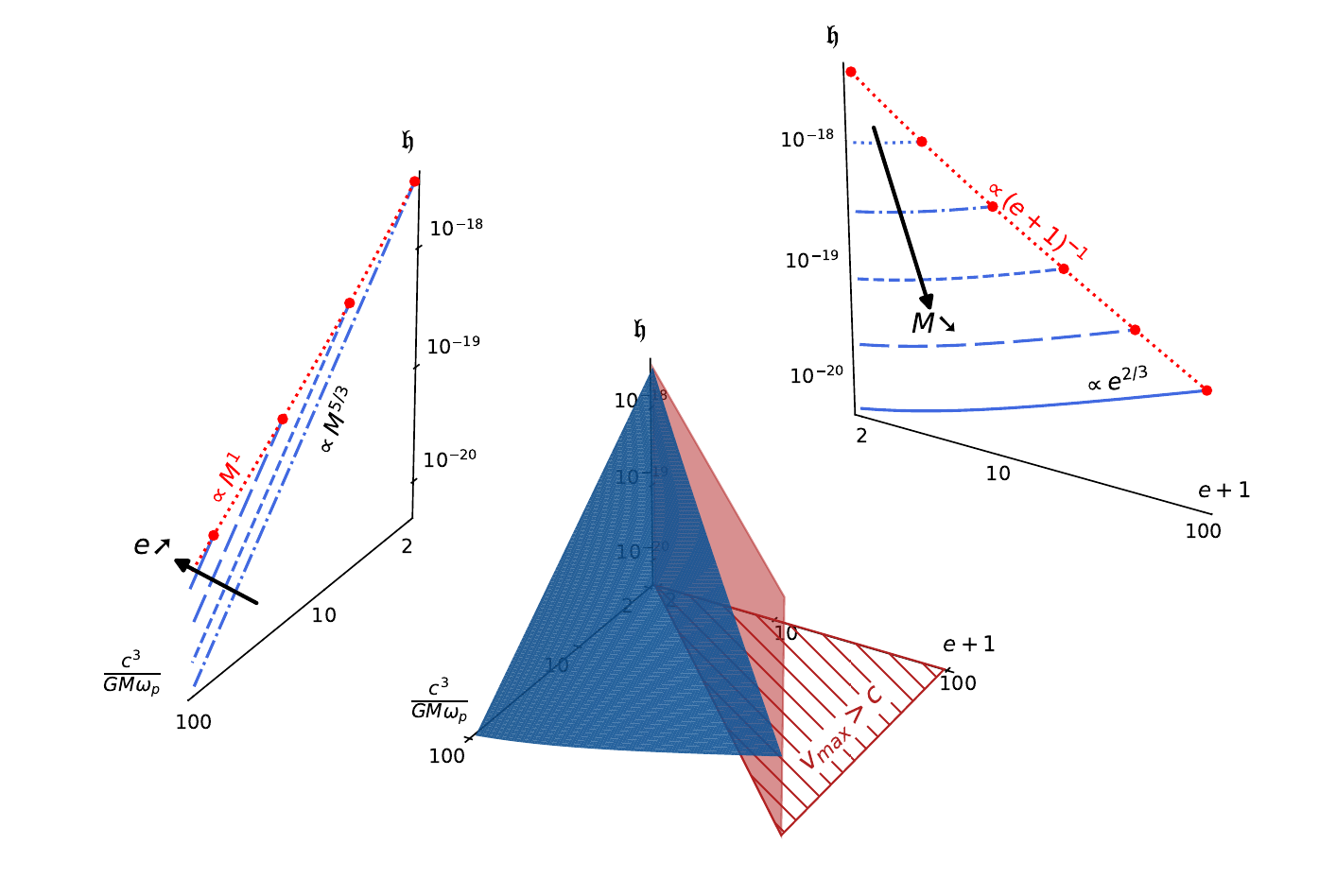}
    \caption{Shape of $\mathfrak{h}\left(\frac{c^3}{\G M \w_p}, e+1\right)$ and projections onto the $\left(\frac{c^3}{\G M \w_p},\mathfrak{h}\right)$ and $(e+1, \mathfrak{h})$ planes (dimensionless variables that are comparable are used). {\it Center panel:} The region $\flatfrac{c^3}{\G M \w_p} < e+1$ is physically forbidden, cf. Eq.~\eqref{eq:Mcritical-newtonian}. {\it Left panel:} Red dots represent the maximal value $M$ at fixed $e$ under physical constraints; they are projections of the intersection between the blue surface and the boundary. One can notice that $\mathfrak{h}$ decreases at a slower pace with respect to this maximally reachable mass than with $M$. {\it Right panel:} Again, red dots are the projection of the intersection between the blue surface and the boundary. Despite that $\mathfrak{h}$ {\it increases} with $e$ at fixed $M$, one observes that the maximal $\mathfrak{h}$ value attainable for a given $M$ actually {\it decreases} with $e$, like $1/(e+1)$. This explains how the strain can simultaneously grows with $e$ yet reaches its global maximum at $e=1$.}
    \label{fig:big-explanation}
\end{figure*}

\bigskip
\bigskip
A visual illustration summarizing the last three subsections is presented on \figref{fig:big-explanation}. We recover that $\mathfrak{h}$ independently grows with $M^{5/3}$ and approximately with $e^{2/3}$. At a given $e$ (or $M$), the maximal $\mathfrak{h}$ value is always reached at the unique point with such $e$ (or $M$) coordinate that lies on the boundary. Most noticeably, the highest value of $\mathfrak{h}$ for a given mass is a {\it decreasing} function of $e$, going like $\flatfrac{1}{(e+1)}$; it epitomizes the apparent paradox detailed above. This power law also appears on the left panel of the figure when considering fixed eccentricities, in which case the upper value of $\mathfrak{h}$ goes like $M$. 

\subsection{Accounting for signal duration}

Considering strain sensitivities of modern GW detectors, say $\mathfrak{h}\sim 10^{-22}$, Eqs. \eqref{eq:absolute-max-h-fiducial-gigahertz} or \eqref{eq:suboptimal-h-fiducial} look rather promising. However, these results must be tempered by the very brief amount of time during which the signal exhibits an interesting feature. This relates to the total integration time during which a detector can accumulate the power of the burst, which in turn determines the detector's signal-to-noise ratio (SNR) for bursts. Typically, it is of the order of \cite{Moore:2014lga,barrau2023prospects}
\begin{equation}
\label{eq:SNR-general}
    \SNR \propto \mathfrak{h}^2 \sqrt{\Delta t} \mcomma
\end{equation}
with $\Delta t$ defined in section \ref{ssub:GWtime}. Some detectors (called haloscopes) even have $\SNR \propto \mathfrak{h}^2 \Delta t$ in some specific sampling regimes \cite{barrau2023prospects}.
For the sake of generality, we define
\begin{equation}
\label{eq:bigxi}
    \Xi(\kappa, e) = \mathfrak{h}_{p}(\kappa,e)^{\alpha}\times \Delta t(e)^{\beta},\qquad \alpha,\beta>0\mperiod
\end{equation}
It was then explicitely computed that if
\begin{equation}
\label{eq:alphaVSbeta}
\beta \leqslant \frac{2}{3}\alpha
\end{equation}
then {\it all} qualitative behaviors demonstrated in sections \ref{sub:abs-max-strain} and \ref{sub:suboptimal-max-strain} about $\mathfrak{h}_{p}$ remain applicable to $\Xi$, namely: {\it (i)}\ the global maximum of $\Xi$ is located at $e=1$ and $\kappa=\flatfrac{c^3}{2\w_p}$, yet {\it (ii)}\; $\Xi$ is a growing function of $e$ at fixed $\kappa$ (and of $\kappa$ at fixed $e$), and so {\it (iii)}\; $\Xi$ reaches its maximum for a generic mass at some $e>1$. In particular, it applies to both SNR expressions mentioned above.

Finally, two formulas are given below for the reader interested in a more quantitative analysis. In the far past/future of the encounter, when $\ihp \to \mp\arccos(-\frac{1}{e})$ and $\w \to 0$, expanding \eqref{eq:time-expression} and \eqref{eq:def-omega} yields 
\begin{equation}
\label{eq:tandphi-far}
    t \approx \mp \pref{\sqrt{\w_p \w(\ihp)}}\sqrt{\frac{e+1}{e-1}} \approx \pref{\w_p (\ihp_\infty - \ihp)}\frac{e+1}{e-1} \mcomma
\end{equation}
while expanding them near $\ihp \to 0$, $\w\to \w_p$ leads to
\begin{equation}
\label{eq:tandphi-close}
    t \approx \mp \frac{1}{\w_p}\sqrt{1+\frac{1}{e}}\sqrt{1-\frac{\w(\ihp)}{\w_p}} \approx \frac{\ihp}{\w_p} \mperiod
\end{equation}
The power laws far and close to the encounter are inversely related, and one can thus see that the interesting feature of \figref{fig:hc-hplus-hcross}  -- the burst -- lasts for a very short period of time.

\subsection{Leveraging various hypotheses}
\label{sub:leveraging}

Below are enumerated the primary approximations and hypotheses upon which our results rely, each followed by a discussion about the robustness of our conclusions if this assumption is leveraged.
\begin{enumerate}
    \item Allowing for $m_1 \neq m_2$: all qualitative behaviors will be preserved regardless, because the basic variables only depend on the total mass $M$. Only $\mathfrak{h}$ is sensitive to the mass ratio, but it is a simple growing function of $\mu$. 
    \item Studying propagation in other directions than $\vu{z}$: we believe that the qualitative behavior will be conserved, at least in a vicinity of the $\vu{z}$ axis, unfortunately the expressions for $h_+, h_\cross$ become too intricate (see \cite{Maggiore:2007ulw}) to directly apply the KKT method.
    \item Replacing $\mathfrak{h}$: focusing on $h_+$ rather than $\mathfrak{h}$ leaves results unchanged as these two quantities are equal at $\ihp=\ihp_p=0$. If one is interested in the power $P_p \propto \kappa^{4/3}(e+1)^{-2/3}$, the behavior is even simpler to grasp: $P_p$ always decreases (resp. increases) with $e$ (resp. $\kappa$), regardless of whether they are free or related by $\kappa(e+1) = \frac{c^3}{\w_p}$, and thus the maximal $P_p$ is always reached on the parabolic orbit. Finally, the case of $h_\cross$ is harder to examine as its maximum is reached at a non-trivial angle, which we derive in appendix \ref{appx:-hcross}.
    \item Considering the signal emitted from other angles than $\ihp = 0$: if one is interested in the signal observable not only at the peak but at any point of the trajectory, our bounds on observable mass ranges like \eqref{eq:Mcritical} would remain valid, since $\w(\ihp) \leqslant \w_p\ \forall\ihp$. They could however be refined further.
     \item Accounting for dissipation effects: as GWs radiate energy away from their source, the orbit eccentricity must be slightly reduced during the encounter, see Eq.~\eqref{eq:def-energy} or Ref.~\cite{caldarola2023effects}. Hence, strictly speaking, an incoming parabolic orbit of $\cal{E}=0$ will necessarily end up with $\cal{E}<0$ and form a bounded system. We can estimate the effect of this loss on near-to-parabolic orbits as follows. Assuming $e=1+\eps,\ \eps\ll 1$, the lost energy reads
     \begin{align}
         \Delta \cal{E} &= \int_{-\ihpinf}^{\ihpinf}P(\ihp)\frac{\dint\ihp}{\dot{\ihp}} \\
         &\approx \frac{170\pi}{3}\G \mu^2 \frac{\kappa^{5/2}}{\ell^{7/2}} \mperiod
     \end{align}
     Comparing this value to the total kinetic energy\footnote{Comparing directly $\Delta \cal{E}$ to $\cal{E}$ is irrelevant because $\cal{E}$ is defined up to a constant, so the fact that $\cal{E}=0$ for a parabola gives no information on the physical energy acquired by the system. On the other hand, $\cal{E}_{c,p}\underset{e\to 1}{\approx}\cal{E}_{c,p}-\cal{E}_{c,\infty}$ is an energy variation and therefore possesses a physical meaning.} of the bodies at the periapsis $\cal{E}_c =\frac{1}{2}\mu\vmax^2$ yields
     \begin{align}
         \frac{\Delta\cal{E}}{\cal{E}_{c,p}} &\approx \frac{85}{8\sqrt[3]{2} c^5} \G\mu \kappa^{2/3}\w_p^{5/3}\\
         &\leqslant\frac{85}{32\sqrt[3]{2}}\left(\frac{\kappa\w_p}{c^3}\right)^{5/3} \\
         &\leqslant \frac{85}{128} \approx 0,66 \qq{using \eqref{eq:Mcritical}.}
     \end{align}
     In extreme scenarios, up to two thirds of the energy of the bodies can be dissipated into GWs! To prevent our previously mentioned results from being invalidated, we need to assume that $e$ is referring to the {\it outgoing} eccentricity $e_f$. Since it is the lowest eccentricity of the orbit, $e_f > 1$ indeed ensures that the trajectory will never become bounded. In this case, our bounds remain true but become largely optimistic and could be refined into stronger inequalities.
\end{enumerate}

\section{Conclusion}
\label{sec:conclu}

This work was dedicated to provide a step-by-step approach to the study of GWs emitted on hyperbolic encounters, as many basic questions arising in such situations are, to the best of our knowledge, scarcely treated in the literature. We first reviewed the standard calculation of the GW signal at lowest order in the multipole expansion, recast with pertinent variables. Then, we showed how the range of masses observable by any detector is intrinsically bounded by a value which solely depends on the operating frequency of the instrument. This bound is very general and relies on a minimal number of hypotheses. We next provided a detailed analysis of the properties of the GW strain as the trajectory eccentricity and the BH total mass evolve. \\

The key feature, somewhat counter-intuitive, is that for any fixed $M$ the parabola generates the {\it minimal} strain among all eccentricities, however the overall {\it maximal} strain observable by a detector is still produced on a parabolic orbit. This, in a nutshell, stems from the physical impossibility of high masses on highly eccentric trajectories to produce high-frequency GWs, without exceeding the speed of light or collapsing onto one-another. Interestingly, this result is highly insensitive to the details of signal-to-noise ratio expression associated with a given type of detector, c.f. equation \eqref{eq:bigxi} and the text below. \\

Furthermore, the potentially large energy loss computed in the last section suggests, as a future study, to consider deviations of the orbit from the Newtonian trajectory, as it has been considered in e.g. \cite{caldarola2023effects}.

In a nutshell, the main purpose of this work was to untangle the already quite subtle situation in a naive Newtonian approximation. This is a mandatory first step as the full picture was missing, even in this elementary framework. Obviously, more refined calculations in the post-Minkowskian program, possibly using amplitude techniques -- where competing factors of Planck’s constant combine into purely classical observables -- will clarify further the situation in the future. It will be paramount to check whether the main conclusions remain correct -- especially about optimal trajectories, and what are the deep underlying reasons invalidating them if it were to be the case.

\bigskip

The early motivation of this work was to challenge the possibility that high-frequency GW detectors (operating around the $\giga\hertz$) could be sensitive to such transient events. While the current article retains its full generality, this new prospect will be thoroughly investigated in a forthcoming note specifically focusing on the high-frequency regime.

\bigskip
\begin{center}
   \small \bf ACKNOWLEDGMENTS
\end{center}

The authors thank Juan Garc\'ia-Bellido for valuable ideas and discussions.

\bigskip
\appendix
\section{Maximum value of \intitle{$h_\cross$}}
\label{appx:-hcross}

This work focused on the typical strain $\mathfrak{h}$, which reaches its maximal value at the peak of the hyperbola, $\ihp=0$. Given \eqref{eq:hplus-value} and \eqref{eq:hcross-value} this is also true for $h_+$ but not $h_\cross$. Let us recall its expression, namely
\begin{equation}
    h_\cross(\ihp) = C g(\ihp), \quad g(\ihp)\equiv \frac{\sin{\ihp}\left(3e +4\cos{\ihp} +e\cos{2\ihp}\right)}{(e+1)^{4/3}}
\end{equation}
with $C$ a simple dimensional constant. In this appendix, we prove that (the absolute value of) $h_\cross$ is maximal at
\begin{widetext}
\begin{equation}
\label{eq:max-of-hcross}
    \pm\ihp_* = \pm 2\arctan\left[\left[\rm{Root}\left((e-1)x^3 -5(e-1)x^2 + 5 (e+1)x -(e+1)\right)\right]^{1/2}\right]
\end{equation}
\end{widetext}
where $\rm{Root}(P_e(x))$ is the unique root of $P_e$ comprised in $\interval{]}{0}{\frac{e+1}{e-1}}{[}$. In particular when $e=1$:
\begin{equation}
    \pm\ihp_* = \pm 2\arctan\pref{\sqrt{5}} \mperiod
\end{equation}

We start by solving for $g'(\ihp)=0$ and replacing $\ihp$ by $t\equiv\tan(\frac{\ihp}{2})$, leading to
\begin{equation}
    (e-1)t^6-5(e-1)t^4+5(e+1)t^2 - (e+1)=0
\end{equation}
where a further replacement of $t$ by $x=t^2$ yields \eqref{eq:max-of-hcross}. However we know that $\ihp$ has to be comprised between $\pm\arccos(-\frac{1}{e})$. We shall therefore demonstrate that $P_e(x)\equiv x^3 -5x^2 + 5 \frac{(e+1)}{e-1}x - \frac{(e+1)}{e-1}$ has one and only one root $x_*$ such that $\ihp_*=2\arctan(\sqrt{x_*})$ (or $-\ihp_*$) lies within this interval (the case $e=1$ is immediate, henceforth $e>1$). 

First, we recast this last condition using $\cos(\arctan u) = \frac{1}{\sqrt{1+u^2}}$ combined with $e>1$: $\left|\ihp_*\right| < \arccos(-\frac{1}{e}) \iff 0< x_* < \frac{e+1}{e-1}$. Moreover, we compute $P'_e(x) = 3x^2 -10x  + 5 \frac{(e+1)}{e-1}$,\ $P_e(0) < 0$ and $P_e(\frac{e+1}{e-1}) > 0$.

\paragraph{$e<4$.} In this case $P'_e$ has no real roots, therefore $P_e$ can vanish at most once on $\R$. But we notice that $P_e(0) P_e(\frac{e+1}{e-1}) < 0$ hence $P_e$ must vanish in $\interval{]}{0}{\frac{e+1}{e-1}}{[}$.
\paragraph{$e=4$.} $P'_e$ has a unique (double) root at $5/3$, however $\flatfrac{5}{3}$ is also equal to $\frac{e+1}{e-1}$. Hence  $P_e$ vanishes at most once in $\interval{]}{0}{\frac{e+1}{e-1}}{[}$ and we conclude as above. 
\paragraph{$e> 4$.} In this situation $P'_e$ has two distinct roots $x_{\pm} = \pref{3}\left(5\pm \sqrt{\frac{10(e-4)}{e-1}}\right)$. Using $e>4$, some algebra shows $x_+ > \frac{e+1}{e-1} > x_- > 0$, implying $P_e$ vanishes at most twice in $\interval{]}{0}{\frac{e+1}{e-1}}{[}$. Let us assume $P_e$ has two distinct roots $x_1 < x_2$ in this interval. Then $P'_e$ has to vanish within $\interval{]}{x_1}{x_2}{[}$ but its root is unique, therefore $x_-$ must belong to $\interval{]}{x_1}{x_2}{[}$. However the fact that $P_e$ vanishes an {\it even} number of times together with $P_e(0)$ and $P_e(\frac{e+1}{e-1})$ having opposite signs implies that one root of $P_e$ must also be a root of its derivative, thus $x_- = x_1$ or $x_- = x_2$, which contradicts the previous statement. It follows that $P_e$ vanishes an odd number of times on $\interval{]}{0}{\frac{e+1}{e-1}}{[}$ which is no greater than two, i.e. exactly once.

\bibliography{references}

\end{document}